\documentclass[footinbib,a4paper,aps,prx,superscriptaddress,reprint,twocolumn,preprintnumbers,amsmath,amssymb,nobalancelastpage,10pt]{revtex4-1}
\usepackage{amsmath}
\usepackage{verbatim}
\usepackage{graphicx}
\usepackage{color}
\usepackage{tikz}
\usepackage{siunitx}
\usepackage{physics}
\usepackage{subfigure}
\usepackage{float}

\usepackage{mathptmx}
\usepackage{amssymb}
\usepackage{amsmath}
\usepackage{amsfonts}
\usepackage{array}

\usepackage{bm}
\usepackage{xcolor}
\usepackage{braket}

\definecolor{mygrey}{gray}{0.35}
\definecolor{myblue}{rgb}{0.2,0.2,0.8}
\definecolor{myzard}{cmyk}{0,0,0.05,0}
\definecolor{mywhite}{rgb}{1,1,1}
\definecolor{myred}{rgb}{1,0.,0.3}

\usepackage[colorlinks=true,citecolor=myblue,linkcolor=myred]{hyperref}

 \def\ee{\mathord{\rm e}}
 
 \def\ii{\mathord{\rm i}}

\def\half{\textstyle\frac{1}{2}}

\renewcommand{\ii}{{\rm i}}
\renewcommand{\ee}{{\rm e}}
\def\beq{\begin{equation}}
\def\eeq{\end{equation}}

\def\barray{\begin{eqnarray}}
\def\earray{\end{eqnarray}}

\begin{document}


\title{Dynamical solitons and boson fractionalization in cold-atom topological insulators}


\author{D. Gonz\'{a}lez-Cuadra}\email{daniel.gonzalez@icfo.eu}
\affiliation{ICFO - Institut de Ci\`encies Fot\`oniques, The Barcelona Institute of Science and Technology, Av. Carl Friedrich Gauss 3, 08860 Castelldefels (Barcelona), Spain}

\author{A. Dauphin}
\affiliation{ICFO - Institut de Ci\`encies Fot\`oniques, The Barcelona Institute of Science and Technology, Av. Carl Friedrich Gauss 3, 08860 Castelldefels (Barcelona), Spain}

\author{P. R. Grzybowski}
\affiliation{Faculty of Physics, Adam Mickiewicz University, Umultowska 85, 61-614 Pozna{\'n}, Poland}
\author{M. Lewenstein}
\affiliation{ICFO - Institut de Ci\`encies Fot\`oniques, The Barcelona Institute of Science and Technology, Av. Carl Friedrich Gauss 3, 08860 Castelldefels (Barcelona), Spain} 
\affiliation{ICREA, Lluis Companys 23, 08010 Barcelona, Spain}

\author{A. Bermudez}
\affiliation{Departamento de F\'{i}sica Te\'{o}rica, Universidad Complutense, 28040 Madrid, Spain}

\begin{abstract}
We study the $\mathbb{Z}_2$ Bose-Hubbard model at incommensurate densities, which describes a one-dimensional system of interacting bosons whose tunneling is dressed by a dynamical $\mathbb{Z}_2$ field. At commensurate densities, the model is known to host intertwined topological phases, where long-range order coexists with non-trivial topological properties. This interplay between spontaneous symmetry breaking (SSB) and topological symmetry protection gives rise to interesting fractional topological phenomena when the system is doped to certain incommensurate fillings. In particular,  we hereby show how topological defects in the $\mathbb{Z}_2$ field can appear in the ground state, connecting different SSB sectors. These defects are dynamical and can travel through the lattice carrying both a topological charge and a fractional particle number. In the hardcore limit, this phenomenon can be understood through a bulk-defect correspondence. Using a pumping argument, we show that it survives also for finite interactions, demonstrating how boson fractionalization can occur in strongly-correlated bosonic systems, the main ingredients of which have already been realized in cold-atom experiments. 
\end{abstract}

\maketitle

The  recent progress   in the field of atomic, molecular and optical (AMO) physics  has established a new paradigm in our understanding of {\it quantum matter}: it is nowadays possible to experiment with  systems of many particles in a pristine and highly-controllable environment. In contrast to solid-state materials, the properties of AMO quantum matter are controlled and probed at the single-particle level, yielding a unique toolbox to explore collective quantum-mechanical effects~\cite{doi:10.1080/00018730701223200}. Historically, systems of ultracold bosonic atoms have played a fundamental role in this regard, allowing for the first experimental demonstration of the condensation predicted by  S. N. Bose and A. Einstein ~\cite{Cornell_2002,Ketterle_2002}.  Additionally, R. Feynman's dream of  a quantum simulator~\cite{Feynman_1982}, i.e. a  controllable quantum  device that can be used as  an efficient alternative to  numerical simulations of quantum many-body problems with classical computers, was also  accomplished for the first time in systems of ultra-cold bosons~\cite{Greiner2002}. This trend has continued in subsequent  years with, among other things, the experimental demonstration  of static gauge fields and  topological phases of matter~\cite{Lin_2009,PhysRevLett.111.185301,PhysRevLett.111.185302,Jotzu2014,Flaschner1091}.

To exploit these quantum simulators in condensed matter or high-energy physics, one typically focuses on AMO systems of fermionic atoms or Bose-Fermi mixtures~\cite{Bloch_2008} and targets  specific models of those disciplines~\cite{doi:10.1080/00018730701223200,1911.00003}. A broader perspective, however, would be to exploit this unique quantum technology to synthesize new forms  of  quantum matter which, while  capturing the essence of  exotic  phenomena in these two disciplines, do so in an entirely different context and via distinct microscopic models that are genuine to  AMO setups. Clearly,  experiments with ultracold bosonic atoms are the key to this endeavor, as no other setup with many interacting bosons can rival with the control and flexibility characteristic of AMO setups. Specifically, ultracold bosons have been used as  quantum simulators of topological insulators, such as the Su-Schrieffer-Heeger model~\cite{Atala_2013,Meier2016,Meier2018}, or the Hofstadter model~\cite{PhysRevLett.111.185301,PhysRevLett.111.185302}, which captures the essence~\cite{Aidelsburger_2014} of the integer quantum Hall effect~\cite{PhysRevLett.45.494} despite consisting of interacting bosons. A current challenge is the quantum simulation of strongly-correlated topological insulators and lattice field theories of matter coupled to dynamical gauge fields.

The former can lead to fascinating physics such as anyons or charge fractionalization~\cite{Laughlin_1999}. Specifically, the fractionalization of electric charge is a groundbreaking phenomenon in  condensed matter, which plays a crucial role in quantum Hall samples. Historically, however, charge fractionalization was first considered in   high-energy physics through  a relativistic quantum field theory of fermions coupled to a solitonic background \cite{relativistic_solitons}. The soliton, which is a  topological excitation that cannot be deformed into the groundstate at a finite energy cost, polarizes the fermionic vacuum  in such a way  that quasi-particles with fractional charge appear in the spectrum. This general phenomenon also takes place in systems where fermions are coupled to phonons \cite{ssh} or to an optical waveguide \cite{Fraser_2019}. In this work, we address the following questions: {\it (i)} could one observe soliton-induced fractionalization  in models genuine to cold-atomic setups?, and {\it (ii)} do solitons lead to charge fractionalization of not only fermionic particles, but also bosonic ones?. A priori, since bosons tend to condense in the lowest energy level, one would naively expect them to be insensitive to the solitonic excitations at higher  energies. Nonetheless, as shown below,  the interplay of strong correlations, topology, and spontaneous symmetry breaking, may cooperate to allow for this exotic effect to occur.

In particular, we study a one-dimensional bosonic system described by the $\mathbb{Z}_2$ Bose-Hubbard model \cite{z2_bhm_1,z2_bhm_2,z2_bhm_3}. We show how, for incommensurate densities, the ground state presents  configurations with topological defects. In particular, a soliton-antisoliton pair with $\mathbb{Z}_2$ charges appear for each extra boson doped above half filling. The bosons get thus fractionalized, and each toplogical defect carries half a boson. Such composite quasiparticles can travel through the system unless externally pinned. Using a local topological invariant, we show how the defects interpolate between regions with different bulk topology, linking it to the presence of fractional particles through a bulk-defect correspondence in the hardcore limit. Finally, we generalize the connection to softcore bosons through a pumping argument, where quantized transport takes place between topological defects.

\begin{figure}[t]
  \centering
  \includegraphics[width=0.9\linewidth]{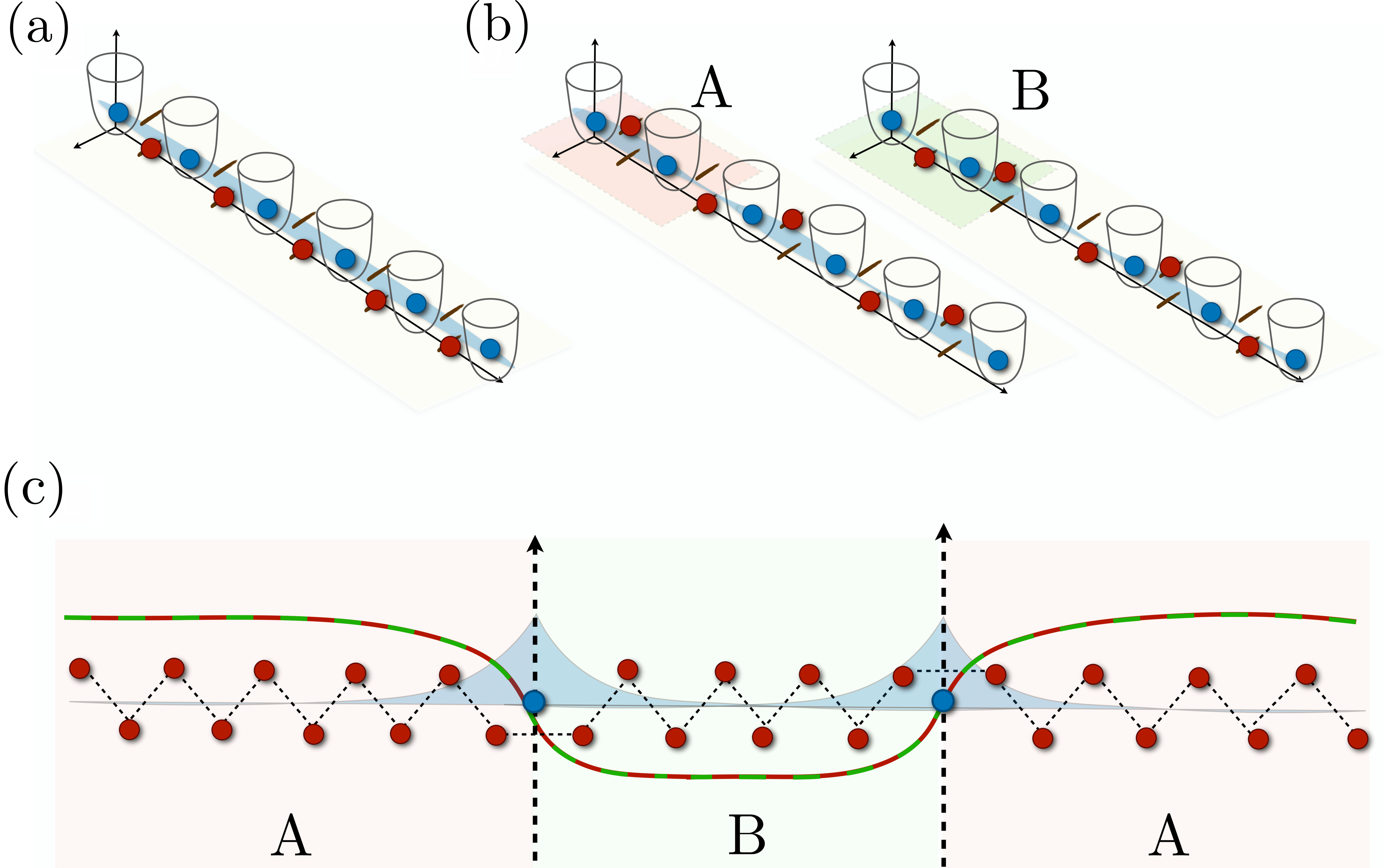}
\caption{\label{fig:scheme} {\bf Bosonic Peierls transition and topological defects:} In the figure, blue and red spheres represent, respectively, bosonic particles tunneling between sites and $\mathbb{Z}_2$ fields located on the bonds. {\bf (a)} At half filling, the $\mathbb{Z}_2$ fields are polarized for weak Hubbard interactions, but order anti-ferromagnetically if the interactions are strong enough, according to two degenerate patterns {\bf (b)}. The SSB drives the bosons from a quasi-superfluid to a BOW phase. Additionally, one of the SSB sectors (B) hosts a SPT phase. {\bf (c)} Extra bosons create pairs of topological defects in the ground state, interpolating between the different SSB sectors and hosting fractionalized particles.}
\end{figure}

\vspace{0.5ex}
{\it The model.--} We consider a lattice field theory of bosons sitting on the vertices of a 1D chain $b_i^{\phantom{\dagger}},b^{\dagger}_i$, where they interact locally with strength $U$,  and get coupled to Ising spins  residing at the lattice links $\sigma^x_{i,i+1},\sigma^z_{i,i+1}$. The spins dress the bosonic bare tunnelling $t$ with a strength $\alpha$, and are subjected to local transverse $\beta$ and longitudinal $\Delta$ fields. This spin-boson lattice field theory is governed by the Hamiltonian
\beq
\label{eq:model}
\begin{split}
H =& -\sum_i \!\left( b_i^\dagger(t+\alpha\sigma^z_{i,i+1})b_{i+1}^{\phantom{\dagger}}+\text{H.c.}\!\right) + \frac{U}{2}\sum_i b^{\dagger}_ib^{\dagger}_ib_i^{\phantom{\dagger}}b_i^{\phantom{\dagger}}\\
&+\frac{\Delta}{2}\sum_i\sigma^z_{i,i+1}+\beta\sum_i\sigma^x_{i,i+1},
\end{split}
\eeq
which is directly motivated by recent experimental progress on Floquet-engineered ultra-cold bosons in optical lattices \cite{dd_fields_eth,z2LGT_2}. This model reminds of a $\mathbb{Z}_2$ lattice gauge theory~\cite{RevModPhys.51.659}, where matter not only interacts  through the $\mathbb{Z}_2$  Ising fields,  but also locally through a 4-body term. The main differences are that the local $\mathbb{Z}_2$ gauge symmetry is explicitly broken in our model, and that we deal with bosonic rather than fermionic matter. We note, however, that  working with fermionic atoms and implementing this gauge symmetry  pose a number of constraints~\cite{Bermudez_2015,PhysRevA.95.053608,Barbieroeaav7444} that complicate the experiments~\cite{dd_fields_eth,z2LGT_2}. Therefore, it is interesting to dispense with them, trying to elucidate if this spin-boson lattice field theory can lead to  genuine quantum matter, hosting {\it  topological fractionalized bosons}.

\vspace{0.5ex}
{\it Groundstate Ising solitons.--} The above Hamiltonian, coined the $\mathbb{Z}_2$ Bose-Hubbard model, has  been studied at  commensurate fillings \cite{z2_bhm_2,z2_bhm_3}. Despite the lack of a Fermi surface, when the Hubbard interactions $U$ are  sufficiently strong, this model displays a spin-boson version of the Peierls instability of  1D electron-phonon systems \cite{peierls}. The Ising spins  spontaneously break the global $\mathbb{Z}_2$ symmetry adopting various magnetic orderings while the bosons, instead of condensing (Fig.~\ref{fig:scheme}{\bf (a)}),  form a  bond order wave (BOW) (Fig.~\ref{fig:scheme}{\bf (b)}). The later  can be understood as an {\it intertwined topological phase},  which simultaneously displays both spontaneous symmetry breaking (SSB) with a local order parameter, and symmetry-protected topological (SPT) features characterized by topological invariants.

Since there are two different SSB configurations for the Ising spins, called A and B in Fig.~\ref{fig:scheme}{\bf (b)}, one can  envisage situations where the spins interpolate between  them forming a soliton (Fig.~\ref{fig:scheme}{\bf (c)}). These finite-energy excitations can be created dynamically by crossing the Peierls transition in a finite time. In this work, we  show that such Ising solitons may also appear in the groundstate for incommensurate fillings  where the system is doped with extra bosons. To analyze this situation, we perform DMRG simulations based on matrix product states (MPS) \cite{tenpy}. In the following, we fix the bond dimension to $D = 100$ and the maximum number of bosons per site to $n_0 = 2$, which is sufficient for strong interactions and low densities \cite{z2_bhm_1}. We also fix the parameter $\alpha = 0.5t$.

\begin{figure}[t]
\centering
\includegraphics[width=1.0\linewidth]{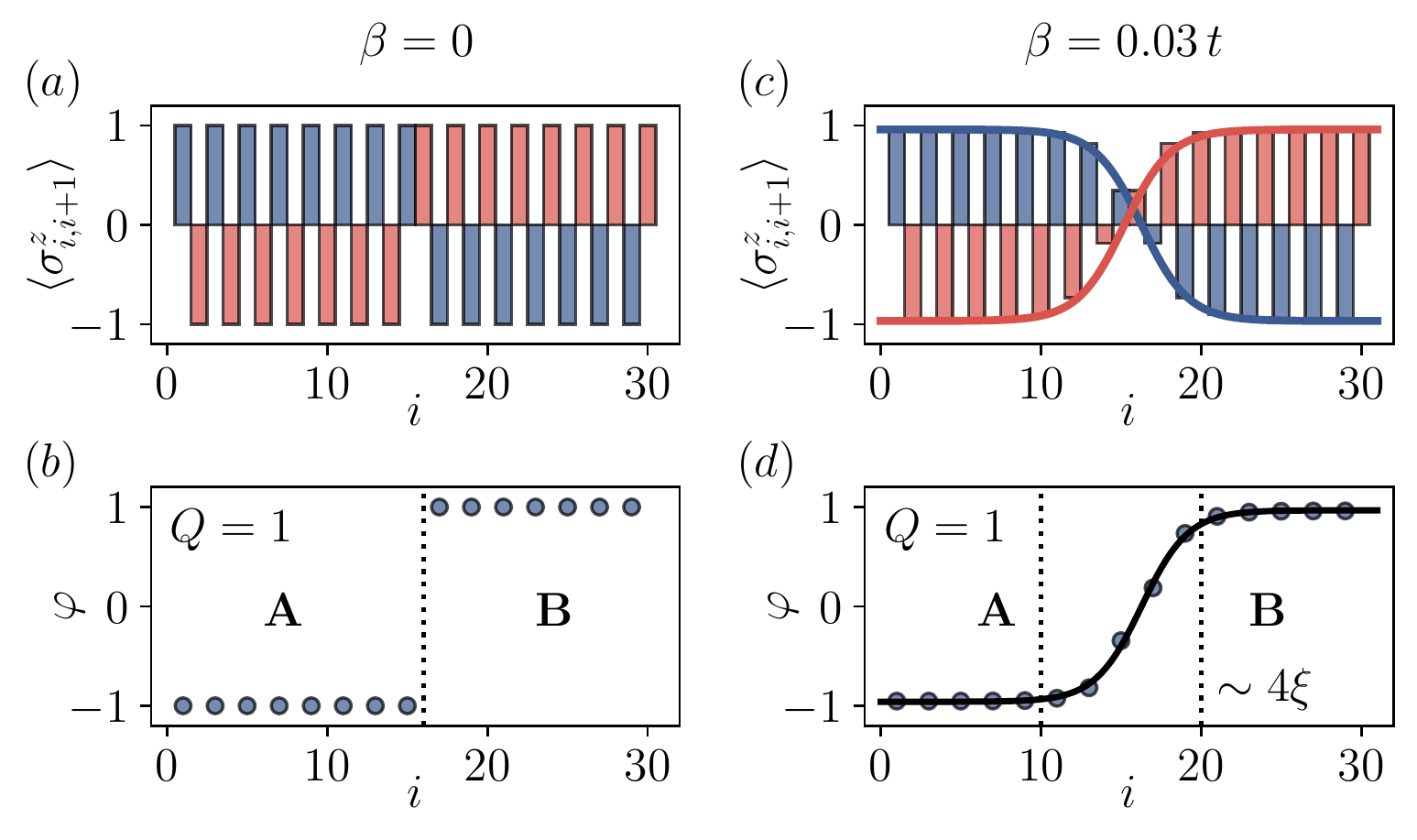}
\caption{\label{fig:ising_soliton} \textbf{Ising topological defects: }
We show the ground state configuration for a chain of $L = 31$ sites and $N = 16$ particles, for $U=10t$, $\Delta = 0.80t$ and different values of $\beta$, where a topological defect appear for the dimmerized pattern of the $\mathbb{Z}_2$ fields. {\bf (a)} and {\bf (b)} show the magnetization $\langle \sigma^z_{i, i+1} \rangle$ and the order parameter $\varphi_j$ for $\beta = 0$, respectively, where the defect is a domain wall with topological charge $Q = 1$ \eqref{eq:top_charge}. {\bf (c)-(d)} Analogous topological defect for  $\beta = 0.03t$, where quantum fluctuations broaden the defect, leading  to a soliton of finite width $\xi$, that can be accurately fitted to Eq.~\eqref{eq:Z2_profile}.}
\end{figure}

In this case, the SSB is characterized by a N\'eel ordering $\langle \sigma_{i,i+1}^z\rangle=(-1)^i\varphi(i)$, 
where  $\varphi(i)$ is a slowly-varying field. The order parameter, defined by the  average $\varphi=\sum_{i}\varphi(i)/N$,  characterizes the two possible SSB sectors $ \varphi=\pm 1$, whereas  solitons correspond to a scalar field $\varphi(i)$  interpolating between these  sectors. As shown in Fig.~\ref{fig:ising_soliton},  solitons present the same  profile as  kinks in  $\lambda\varphi^4$ relativistic field theories~\cite{PhysRevD.10.4130}, namely
 \beq
 \label{eq:Z2_profile}
 \varphi(i)=\tanh \left(\frac{i-i_{\rm p}}{\xi}\right),
 \eeq 
 where $i_{\rm P}$ is the soliton centre and  $\xi$  its width. By analogy with the   scalar quantum field theory~\cite{Rajaraman:1982is}, the topological charge $Q=\half\!\int\!{\rm d}x\partial_x\varphi(x)$ can be evaluated by a finite difference
 \beq
 \label{eq:top_charge}
 Q= \half\left(\varphi(i_{\rm p}+r)-\varphi(i_{\rm p}-r)\right),
 \eeq
at points  well separated from the soliton center $r/\xi\sim\mathcal{O}(N)$.

Figures~\ref{fig:ising_soliton} {\bf (a)}-{\bf(b)} show the results for the $\beta=0$ limit, where the Ising spins have no quantum fluctuations, and the solitons reduce to static domain walls, the center of which can be found anywhere within the lattice.  A topological charge of $Q=+1$ can be directly read from  the soliton of Fig.~\ref{fig:ising_soliton}{\bf(b)}.
We remark two differences with respect to    relativistic field-theory solitons: our solitons {\it (i)}  appear directly in the groundstate, and {\it (ii)}  are not free to move due to Peierls-Nabarro barriers caused by the lack of a continuous translational invariance~\cite{Peierls_nabarro_1,Peierls_nabarro_2}.

As $\beta>0$ is switched on, the Ising spins are no longer classical discrete variables,  but become dynamical  fields with quantum-mechanical fluctuations. A direct consequence is that they can tunnel through the barriers and delocalize over the chain. To benchmark  the prediction~\eqref{eq:Z2_profile}, we introduce a pinning mechanism by raising the transverse field  $\beta\to\beta_0=\beta(1+\epsilon)$ at  two  consecutive bonds  surrounding  the desired pinning centre $i_{\rm p}$. In Figs.~\ref{fig:ising_soliton} {\bf (c)}-{\bf(d)}, we show that the effect of quantum fluctuations is to widen the extent of the soliton,  such that the interpolating region has now a non-zero width $\xi>0$ that can be extracted by a fit of the corresponding scalar field $\varphi(i)$ to Eq.~\eqref{eq:Z2_profile}. In Ref.~\cite{long_paper}, we present a quantitative study of the aforementioned Peierls-Nabarro barriers, and the quantum-induced widening of the soliton, paying special attention to the role of the repulsive Hubbard interactions that controls the back-action of the bosonic matter on the Ising solitons. Moreover, we  explore other fillings that lead fo solitons with higher topological charges which, nonetheless, still display clear analogies with the  $\lambda\varphi^4$  kinks~\cite{PhysRevD.10.4130}.

\vspace{0.5ex}
{\it Fractionalization of bosons.--} The fact that the Ising solitons are not restricted to finite-energy excitations,  as typically occurs  in relativistic quantum field theories, but appear instead in the groundstate is crucial if one aims at finding a bosonic version of charge fractionalization. The bosons, which tend to condense in the lowest-possible energy level, will do so forming a bond-ordered wave that can lead to fractionally-charged quasi-particles. An unambiguous manifestation can be found by doping the half-filled system with a single   particle. To accommodate for this  particle,   an Ising soliton/anti-soliton pair is created, each  hosting a bound quasi-particle with a fractionalized  number of bosons, i.e. the boson splits into two halves.

\begin{figure}[t]
  \centering
  \includegraphics[width=1.0\linewidth]{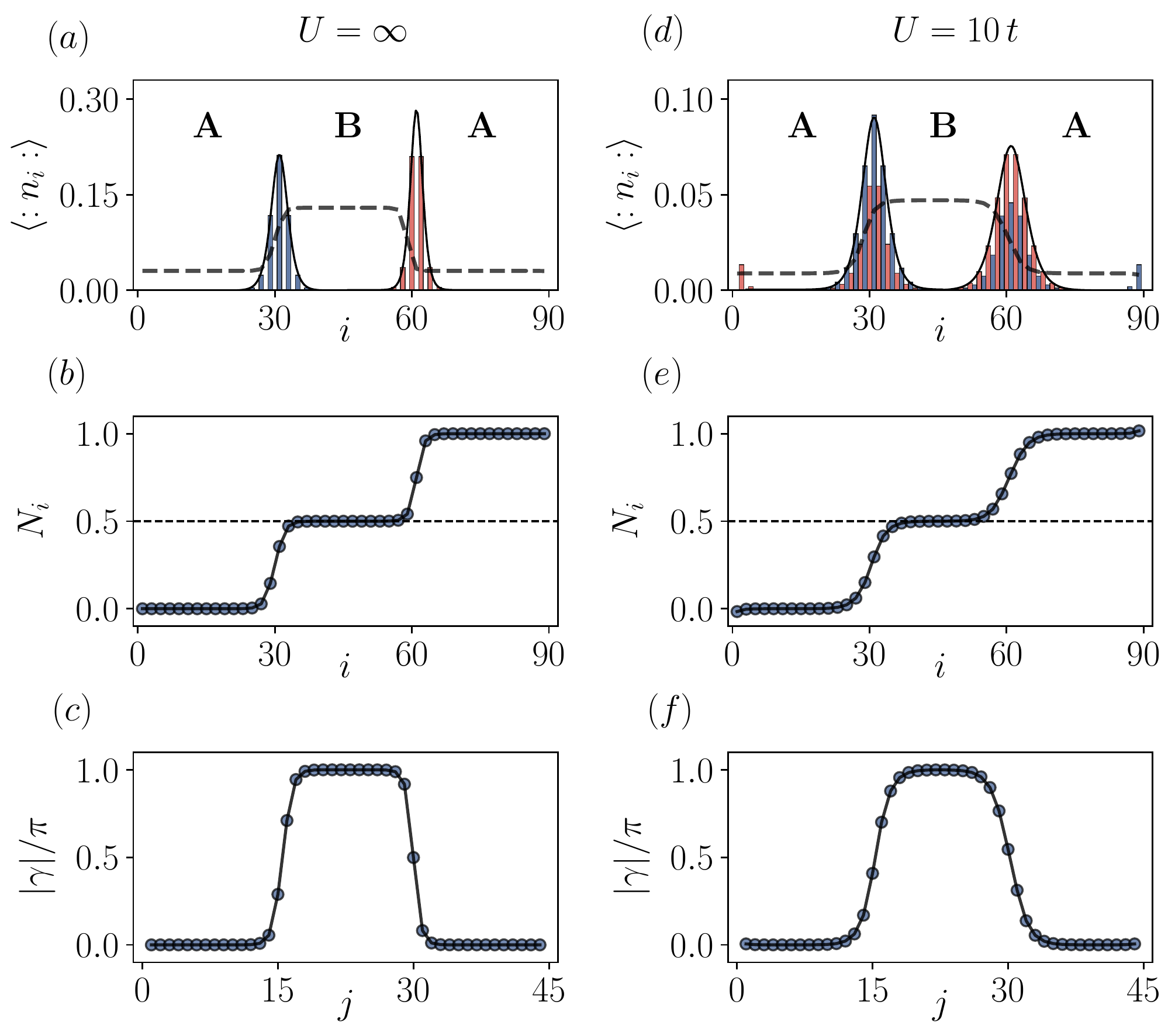}
\caption{\label{fig:spt_dimer} \textbf{Boson fractionalization and local Berry phase: } {\bf (a)} Occupation number $\langle : n_i : \rangle$ for a chain with $L = 90$ sites and $N = 46$ particles in the hardcore limit, for $\Delta = 0.70t$ and $\beta = 0.03t$. We rescaled and superimposed the order parameter $\varphi$ using dashed lines. We observe peaks in the occupation at the defects, localized on different sub-lattices (represented in different colors) due to chiral symmetry. The solid lines correspond to a fit to Eq.~\eqref{eq:Z2_boson_mode}. {\bf (b)} The integrated particle number $N_i$ shows how each peak contains half a particle. {\bf (c)} The local Berry phase $\gamma$ for each unit cell $j$ is quantized to $0$ or $\pi$ in the different SSB sectors, and interpolates between the two around the defects. {\bf (d)} Softcore bosons with $U = 10t$, $\Delta = 0.80t$ and $\beta = 0.02t$, where we still observe peaks in the occupation, associated with a fractionalized boson {\bf (e)}. These are no longer localized in specific sublattices, since chiral symmetry is broken. However, the topological Berry phase is still quantized far from the defects {\bf (f)} since inversion symmetry is still preserved.}
\end{figure}

This fractionalization mechanism  is  confirmed by the numerical results  presented in Fig.~\ref{fig:spt_dimer}. The scalar field associated to the Ising spins displays the aforementioned soliton-antisoliton  pair for a chain of $N=90$ sites and filled with $N_{\rm b}=46$ bosons (see Fig.~\ref{fig:spt_dimer}{\bf (a)} and {\bf (d)}). One clearly sees that there is a density build-up around  the topological defects that follows the superposition of two   profiles  
 \beq
 \label{eq:Z2_boson_mode}
 \langle :n_j:\rangle= \langle n_j\rangle-\frac{1}{2}=\frac{1}{4\xi}\sech^2 \left(\frac{j-j_{\rm p}}{\xi}\right),
 \eeq 
for the even $j = 2i$ or odd $j=2i+1$  sub-lattices, with their corresponding centers $j_{\rm p}$ being fixed by the soliton-antisoliton positions. We note that this expression coincides   with the profile of fermionic  zero-modes for the relativistic Jackiw-Rebbi model~\cite{CB_solitons_GN} where  fractionalization was first predicted~\cite{relativistic_solitons}.  In order to test if the bosons fractionalize, we represent in Fig.~\ref{fig:spt_dimer}{\bf (b)} and {\bf (e)} the integrated density $
N_i=\sum_{j\leq i} \langle :n_j:\rangle $, which shows two  plateaux where the boson change  jumps by  steps  of $1/2$. One can thus interpret that the soliton and antisoliton  bind  half a  boson each, forming two fractionalized quasi-particles. 

In the companion article \cite{long_paper}, we present an in-depth analysis of this fractionalization phenomenon, ruling out the existence of polaron quasi-particles, and showing that the soliton dynamics  is crucial to understand  the appearance of self-assembled  soliton lattices with a periodic arrangement of the fractional charges. Moreover, we
  extend the analysis  to situations where  particles are doped above  other commensurate fillings,  and argue that the larger topological charges of the soliton allows for other fractional densities.

  \vspace{0.5ex}
{\it Many-body topological invariants.--} In this part of the manuscript, we show that the topological characterization of our spin-boson groundstate  is not fully captured by the  charge of the classical scalar field~\eqref{eq:top_charge}. We show that the bosonic sector  may also display a non-zero  local Berry phase, the calculation of which requires a full quantum-mechanical treatment    of the many-body groundstate
\beq
\gamma_i=\int_{0}^{2\pi}\!\!{\rm d}\theta_i\bra{\epsilon_{\rm gs}(\theta_i)}\ii\partial_{\theta_i}\ket{\epsilon_{\rm gs}(\theta_i)},
\eeq 
where $\theta_i$ is the angle that twists the local bosonic tunneling $t\to t\ee^{\ii\theta_i}$ between sites $i$ and 
$i+1$. This local Berry phase generalizes the notion of a many-body Berry phase with twisted boundary conditions~\cite{Niu_1984}, and can be calculated for situations that are not translationally invariant~\cite{hatsugai_2006}, which is crucial in the present  context of  solitonic Ising fields.

In Figs.~\ref{fig:spt_dimer}{\bf (c)} and {\bf (f)}, one can see how the local Berry phase at the links joining two neighbouring unit cells changes by $\Delta \gamma=\pm\pi $ as one traverses the soliton/antisoliton. Therefore, the topological defects  not only carry a topological charge $Q=\pm1$, but they separate  topologically-trivial  regions $\gamma_A=0$ from non-trivial  ones  $\gamma_B=\pi$. We note that the theory of defects in symmetry-protected topological phases of matter~\cite{topological_defects_1, topological_defects_2} relates such inhomogeneous layouts of topological invariants  with the existence of protected  quasi-particles localized at the defects. 

\begin{figure}[t]
\includegraphics[width=1.0\linewidth]{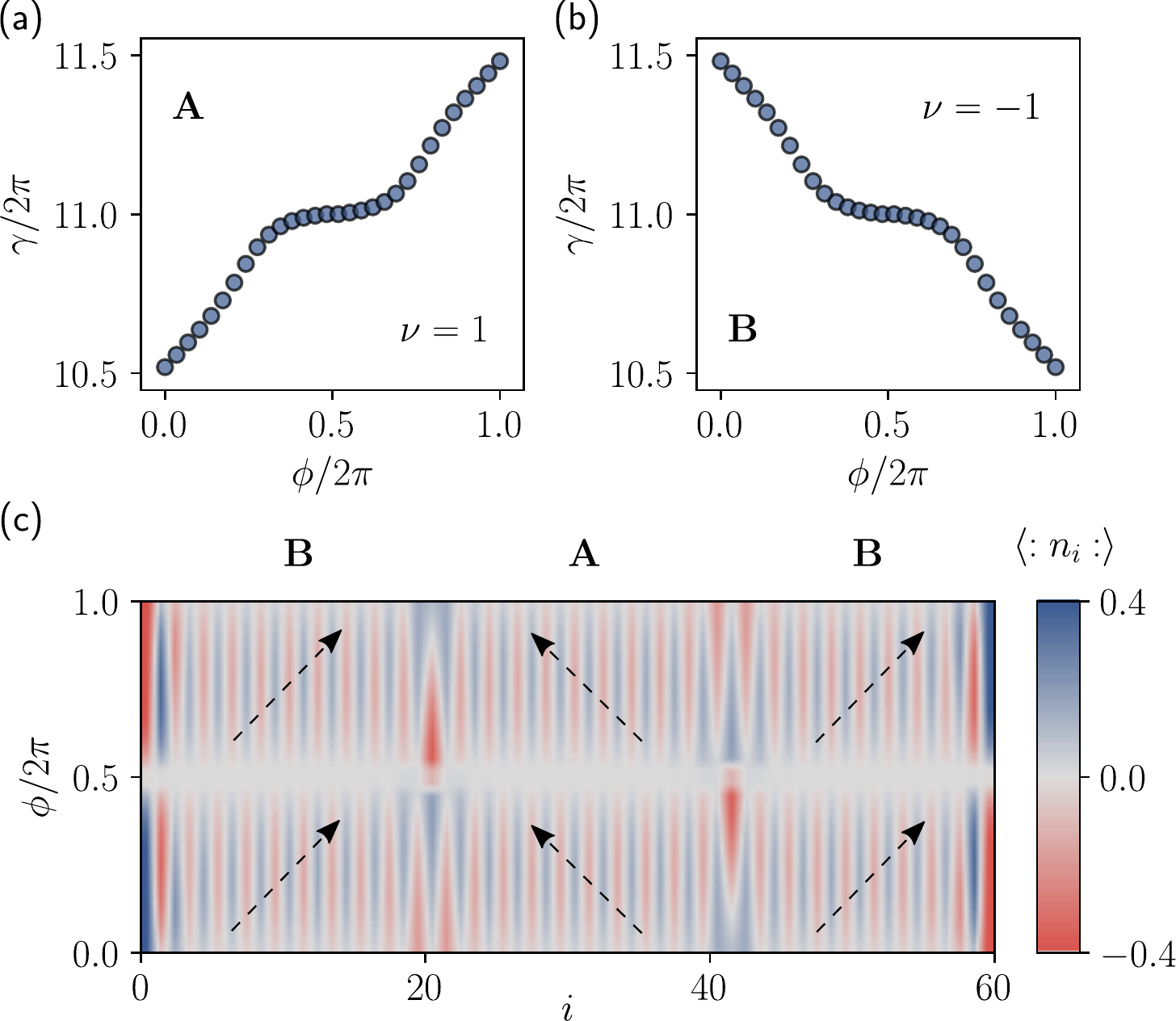}
\caption{\label{fig:chern_entanglement} \textbf{Bosonic pumping between edges and defects: } We show how the Berry phase $\gamma$ changes with the pumping parameter $\phi$, calculated at the middle of a chain with $L=42$ and $N=21$. The change in this quantity \eqref{eq:chern_berry} is associated to the Chern numbers of an extended 2D system, where {\bf (a)} $\nu_{\rm A}=1$ and {\bf (b)} $\nu_{\rm A}=-1$, correspond to the 1D configurations A and B, respectively, at $\phi = 0$. {\bf (c)} Bosonic occupation $\langle : n_i : \rangle$ through the pumping cycle of a system with two domain walls, starting from a BAB configuration at $\phi = 0$, where the topological edge states are visible. Each region transports a quantized charge in the bulk given by $\Delta n = -\nu$.}
\end{figure}

Let us emphasize that this theory, however, deals with fermions and typically assumes an  externally-adjusted  defect that only serves to provide a background for the fermions. To the best of our knowledge, our results show for the first time that analogous effects occur for bosons in a fully-fledged quantum many-body problem where the defects are dynamical solitons with their own quantum fluctuations. Following this connection, we note that in the $U/t\to \infty$ limit, our spin-boson model  has an additional chiral (sub-lattice) symmetry. As confirmed by our results of Figs.~\ref{fig:spt_dimer}{\bf (c)} and {\bf (f)}, the topological characterization remains the same, but the the  quasi-particles localized at the soliton/antisoliton states have support  in just one of the two sub-lattices, such that they are protected  against perturbations that respect this chiral symmetry. Therefore, apart from the inherent robustness of the classical topological solitons, the total defects formed by a soliton and a fractionalized  bosonic quasi-particle are also protected against chiral-preserving perturbations.

  \vspace{0.5ex}
{\it Quantized boson transport between edges and defects.--} It is interesting to note that, away from the $U/t\to \infty$ limit, the fractionalized quasi-particles do not enjoy the additional robustness of SPT arguments. In this section, however, we show that an adiabatic  inter-soliton and edge-soliton transport of bosons through a Thouless pump shows a robust quantization, and that these quasi-particles can be understood as the remnants of higher-dimensional defects that do have this additional robustness. To induce this  pumping, we  drive the transverse and longitudinal Ising fields along a periodic cycle $\Delta\to\Delta_i({\phi})=2(-1)^it\cos\phi$, $\beta\to\beta_i({\phi})=(-1)^it\sin\phi$, where $\phi:0\to2\pi$. 

In fermionic SPT phases, such adiabatic cycles lead to the transport of an integer number of fermions $\Delta n$ across the system, which coincides with the Chern number $\Delta n=\nu$  of an effective 2D system~\cite{niu1985}. Alternatively~\cite{ent_spectrum_berry,ent_spectrum_berry_2}, this  invariant can be obtained from the change of the  Berry phase 
\begin{equation}
\label{eq:chern_berry}
\nu = \frac{1}{2\pi}\int_0^{2\pi}\text{d}\phi\,\partial_\phi \gamma (\phi),
\end{equation}
  which can be calculated in the many-body case using the approach of the previous section. As shown in Fig.~\ref{fig:chern_entanglement}, this pumping mechanism can also be applied to bosons in the presence of dynamical solitons. In this case, as shown in Figs.~\ref{fig:chern_entanglement}{\bf (a)}-{\bf (b)}, the change of the local Berry phase at the middle of the chain yields the Chern numbers $\nu=\pm1$ depending on the SSB sector A/B of the initial state.  By calculating the evolution of the boson density (Fig.~\ref{fig:chern_entanglement}{\bf (c)}), one clearly observes that a single bosonic charge is transferred between the soliton and antisoliton, and between each of them and the closest edge.  Comparing with  Figs.~\ref{fig:chern_entanglement}{\bf (a)}-{\bf (b)},  it becomes clear that this pumping is directly associated to the different  Chern numbers.
  
  As announced above, the fractional bosons bound to the defects can be understood as remnants of the higher-dimensional conducting states that are localized at the interfaces
  separating the 2D regions of different Chern number. In the companion paper \cite{long_paper}, we present a thorough analysis of this pumping mechanism, and show that it is crucial to understand the topological properties of the soliton quasi-particles that appear at other fractional fillings, unveiling a generalized bulk-defect correspondence.

  \vspace{0.5ex}
{\it Conclusions and outlook.--} In this work we showed how boson fractionalization can take place in cold-atomic systems. In particular, we study the ground state of the $\mathbb{Z}_2$ Bose-Hubbard model for incommensurate densities around half filling, where we found composite quasiparticles consisting on dynamical solitons on the $\mathbb{Z}_2$ field together with particles with a fractional bosonic occupation number. We also characterized their properties, including the topological and the fractional charge. Finally, we connected these properties to the topological character of the underlying bulk through a generalized bulk-defect correspondence, where we demonstrated the quantization of inter-soliton transport.

\acknowledgements

This project has received funding from the European Union’s Horizon 2020 research and innovation programme under the Marie Sk\l{}odowska-Curie grant agreement No 665884, the Spanish Ministry MINECO (National Plan 15 Grant: FISICATEAMO No. FIS2016-79508-P, SEVERO OCHOA No. SEV-2015-0522, FPI), European Social Fund, Fundaci\'{o} Cellex, Generalitat de Catalunya (AGAUR Grant No. 2017 SGR 1341, CERCA/Program), ERC AdG NOQIA, EU FEDER, MINECO-EU QUANTERA MAQS, and the National Science Centre, Poland-Symfonia Grant No. 2016/20/W/ST4/00314. A. D. is financed by a Juan de la Cierva fellowship (IJCI-2017-33180). A.B. acknowledges support from the Ram\'on y Cajal program RYC-2016-20066, and CAM/FEDER Project S2018/TCS- 4342 (QUITEMAD-CM).

\bibliographystyle{apsrev4-1}
\bibliography{bibliography}

\end{document}